\documentclass[%
 reprint,
superscriptaddress,
longbibliography,
 amsmath,amssymb,
 aps,
prb,
]{revtex4-1}
\setcitestyle{numbers,square}

\usepackage{braket}
\usepackage{graphicx}
\usepackage{dcolumn}
\usepackage{bm}
\usepackage[normalem]{ulem} 
\usepackage[pdftex,usenames,dvipsnames]{color}
\usepackage{siunitx}
\usepackage[version=4]{mhchem}
\usepackage{amsmath,empheq}
\usepackage{float}
\usepackage[T1]{fontenc}
\usepackage{tgbonum}

\usepackage{xcolor}

\DeclareSIUnit\elementarycharge{\text{\ensuremath{e}}}
\DeclareSIUnit\angstrom{\text {Å}}
\DeclareSIUnit\uc{\text {unit-cell}}

\usepackage{hyperref}
\hypersetup{
    colorlinks,%
    citecolor=blue,%
    linkcolor=blue,%
    urlcolor=blue
}

\newcommand{\orcid}[1]{\href{https://orcid.org/#1}{\includegraphics[width=8pt]{orcid.pdf}}}

\begin{document}

{\fontfamily{qcr}\selectfont
}

\title{Anatomy of torques from orbital Rashba textures:\\ the case of Co/Al interfaces}

\author{A. Pezo}
\author{N. Sebe}
\affiliation{Laboratoire Albert Fert, CNRS, Thales, Universit\'e Paris-Saclay, 91767 Palaiseau, France}
\author{A. Manchon}
\affiliation{Aix-Marseille Université, CNRS, CINaM, Marseille, France}
\author{V. Cros}
\affiliation{Laboratoire Albert Fert, CNRS, Thales, Universit\'e Paris-Saclay, 91767 Palaiseau, France}
\author{H. Jaffrès}
\affiliation{Laboratoire Albert Fert, CNRS, Thales, Universit\'e Paris-Saclay, 91767 Palaiseau, France}
 
\email{armando-arquimedes.pezo-lopez@cnrs-thales.fr,henri.jaffres@cnrs-thales.fr}

\date{\today}

\date{\today}

\begin{abstract}
In the context of orbitronics, the rising of the orbital angular momentum generated at light metal interfaces from orbital textures via orbital Rashba-Edelstein effects nowadays represent extraordinary alternatives to the usual heavy-metal spin-based materials. In the light of very recent experimental results [S. Krishnia \textit{et al.}, Nanoletters  2023, 23, 6785], starting from state-of-the-art density functional theory simulations, we provide theoretical insights into the emergence of very strong orbital torques at the Co/Al interface location a strong orbital Rashba texture. By using linear response theory, we calculate the exerted orbital torque amplitudes, mainly of field-like intraband character, acting onto the ultrathin Co. Moreover, we show that an insertion of a single atomic plane of Pt between Co and Al is enough to suppress the effect which questions about the anatomy of the torque action clearly behaving differently than in the standard way. This work opens new routes to the engineering of spintronic devices. 
\end{abstract}

\maketitle

\noindent
{\it Introduction} - Orbital Hall (OHE) \cite{PhysRevLett.121.086602,pezo_ohe,salemi2022} and orbital Rashba-Edelstein effects (oREE) \textit{via} the orbital angular momentum (OAM)~\cite{Nikolaev2024, PhysRevB.103.L121113,Johansson2021,salemi2021, Johansson2024,pezo2024spinorbitalrashbaeffects,Krishnia2023} have recently emerged as new sources to knob quantum mechanical transport properties that can be potentially disrupting for the future generation of spintronic devices such orbital SOT-MRAM or spintronics THz emitters~\cite{SOT_Roadmap}. In a recent theoretical reference, Nikolaev \textit{et al}~\cite{Nikolaev2024} identified an orbital texture at the interface between the light metals Co and Al and they estimated the enhanced torques on Co resulting from the current-induced orbital Rashba-Edelstein polarization associated with the orbital texture, featuring thus possible novel torque functionalities. Moreover, the observation of long-range orbital generation of orbital accumulation at the surface like Ti~\cite{Choi2023} and Cr~\cite{magneto_Cr_2023} by magneto-optical techniques have largely strengthen the interest for light metals (LM). Those phenomena appear as their spin counterparts, as the spin Hall (SHE) and spin Rashba-Edelstein effect (sREE), offering new mechanisms for data storage and energy-efficient electronics with the important difference lying on the dispensable role of spin-orbit interactions (SOI). SHE relying on heavy metals (HM) as Pt, W or Ta was already proven to realize efficient spin-orbit torque (SOT) functionalities~\cite{Hirsch_1999_SHE, Sinova_2015_revueSHE, Liu_2011_SHESTTFMR, Liu_2012_SHEswitch} also involving the orbital degree of freedom~\cite{oppeneer2025}. The generation of a transversal orbital current resulting from an applied electric field ($\mathcal{E}$), \textit{e.g.} via OHE, has garnered interest because it may interact with local moments over long-range distance~\cite{Hayashi2023,longrangeorbital} thus enabling torques without any HM or even switch the magnetization with higher efficiency with the use of Nb, Ru and Cr LMs~\cite{Gupta2025}. The demonstration of SOT in LM, such as Ti, has revealed the required properties to effectively achieve control of magnetization dynamics~\cite{go_2017_toward, S_Ding_ORE, Go_2021_ORE, S_Ding_OREM}. Additionally, the discovery of a large interfacial inverse oREE, along with recent terahertz (THz) emission using Ni~\cite{Seifert2023,Xu2024NatCom,zhao2025} or CoPt alloys~\cite{Xu2024AdvMat} as orbital sources, shows promise for uncovering new quantum properties in previously uncharted electronic devices.

From fundamentals viewpoint, oREE occurs when the electric field $\mathcal{E}$ induces a non-equilibrium orbital angular momentum (OAM) polarization in materials via local current gradient~\cite{raimondi2025} or at their interfaces lacking inversion symmetry. This feature appears as the counterpart of sREE generating a spin polarization via SOI in systems with broken inversion symmetry~\cite{Rashba_1984, EDELSTEIN_1990, mihai2010current, manchon2015perspectives}. Nonetheless, it is generally admitted that the lack of coupling between OAM and magnetization represents a main difficulty to use them for manipulating a magnet. Still, oREE has the potential to convert OAM into spin and then interact with magnetic moments through SOI, possibly offering new efficient mechanisms for spintronic applications. Recent experiments~\cite{ElHamdi2023} and theories~\cite{Johansson2021,Johansson2024}  have demonstrated orbital-charge conversion in oxide 2-dimensional electron gas, or with topological insulators~\cite{PhysRevResearch.6.043332}. It is anticipated that the latter conversion could be up to five or six times more efficient than the spin-to-charge conversion \textit{e.g.} observed by terahertz spectroscopy~\cite{Rongione2023}. Additionally, the generation of orbital photocurrents at BiAg$_2$ sufaces results in a significant orbital Rashba effect, presenting an excellent opportunity to manipulate electrical currents using the intrinsic orbital momentum of the light~\cite{Mokrousov2024}. Furthermore, the large oREE in synthetic multiferroics accompanying its spin counterpart has been shown to produce substantial values capable of magnetization control~\cite{pezo2024spinorbitalrashbaeffects}. 

\begin{figure}[ht!]
    \centering
    \includegraphics[width=\linewidth]{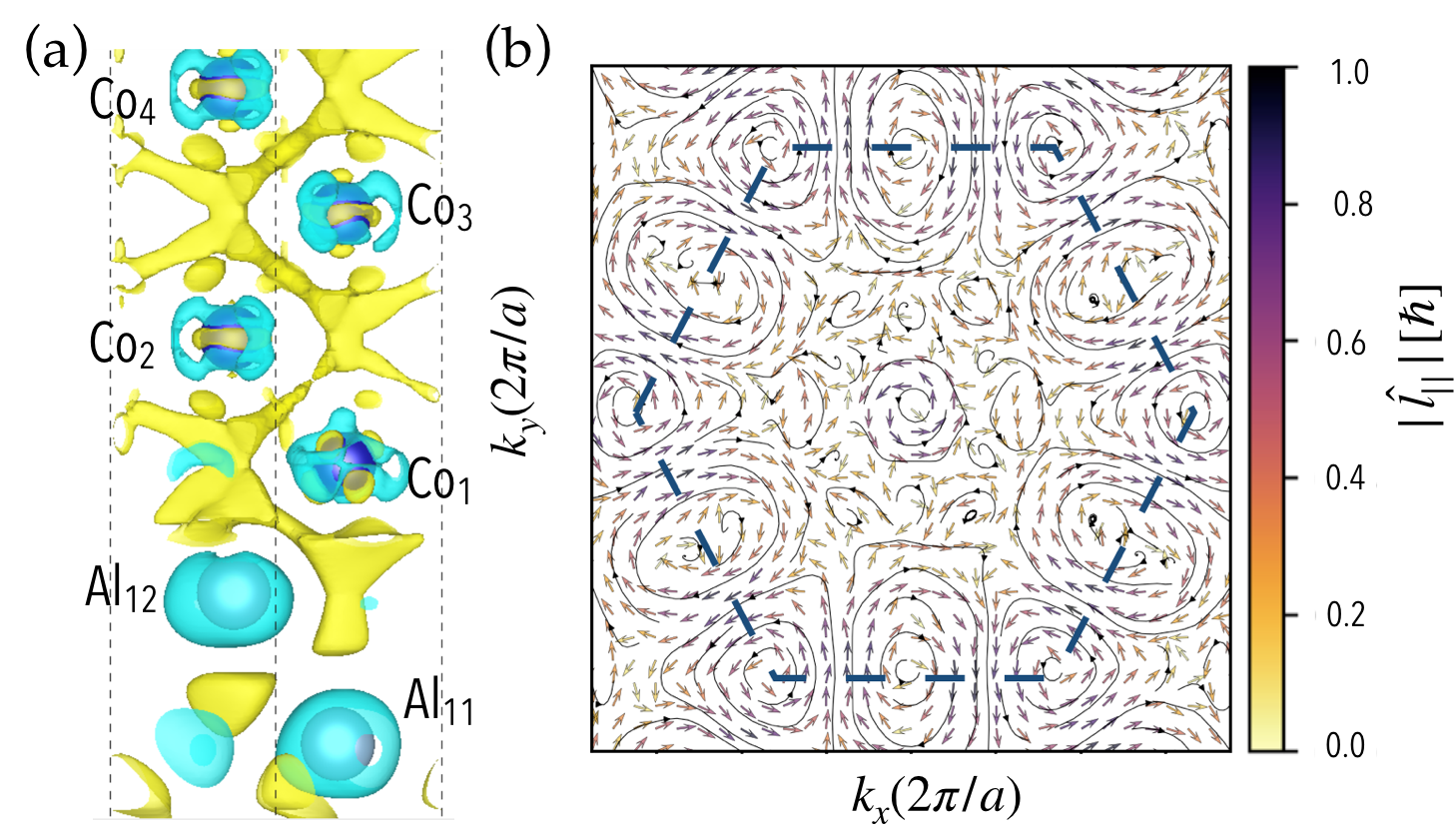}
    \caption{
    Charge distribution at Co(12)|Al(12) interface (a) where the charge difference resides at the interface between Co and Al layers for an isosurface of $2\times 10^{-4}$ e/~\AA$^3$. yellow (cyan) colors means charge accumulation (depletion) regions (b) The orbital texture of hexagonal shape calculated at the Fermi level within the Brillouin zone represented by a dashed line.}
    \label{fig:bands}
\end{figure}

In order to expand the range of materials able to offer spin manipulation, the orbital degree of freedom, ubiquitous in all electronic systems, has thus emerged as an attractive option. Nevertheless, spin and orbital degrees of freedom are intertwined, making it generally challenging to differentiate them. This complexity makes the use of low-Z materials essential~\cite{Go_2020_OTgen} and the choice of the ferromagnet may have a significant quantitative impact on device improvement and design~\cite{Kim_2021_nontrivial}. As recently demonstrated~\cite{Krishnia2023}, and among different Co/LM structures, we presently conclusively demonstrate the existence of the orbital torque (OT) at Co/Al interfaces by means of joint Density Functional Theory (DFT) calculations with Kubo linear response theory; with the benefit here of avoiding orbital transport over long distance. By analyzing the corresponding interface properties largely explored~\cite{Krishnia2025} and subsequent electronic structure, it is concluded, in the present work, that the generated out-of-equilibrium orbital density exerts strong OT on an ultrathin Co magnet demonstrating thus that oREE represents the main contribution contrasting with the effect obtained by placing Pt, a reference SHE layer, sandwiched between Co and Al.\\
\noindent

{\it Co/Al interfaces - band structure -} We have built a slab containing twelve layers of each fcc Al and hcp Co grown onto the same [111] directions~\cite{patel2023} and with Co having the magnetization $\mathbf{M}=M_z$ pointing along the non-periodic direction $\hat{z}$ normal to the layers. The in-plane lattice parameter was fixed at $a=2.94$ \AA~ corresponding to energy minimization (\textcolor{blue}{Supp. Info. SI-I}). 
By using the self-consistent converged ground state, we obtained the electronic structure of Co(12)/Al(12) projected onto the first Co(1) atomic plane in contact with Al(12) where we have signatures of the OREE. This manifests the orbital texture displayed on Fig.~\ref{fig:bands}(b) as a result of the electrostatic field appearing at the Co/Al interface clearly visible in Fig.~\ref{fig:bands}(a) displayed by the charge density difference suggesting a charge accumulation region on the Co interfacial layer. We observe that the orbital texture, the two $\braket{\hat{l}_{x,y}}$ in-plane orbital components calculated onto the Fermi surface, as depicted in Fig.~\ref{fig:bands}(b), manifests a chiral texture of large OAM amplitude reaching 0.6-0.8 $\mu_B$ unlike the Co(12)/Cu(12) 'reference' interfaces showing no specific OAM feature (\textcolor{blue}{SI-II}). The ensemble of these results describing the electronic band structure and OAM properties of Co/Al and Co/Cu systems are in both qualitative and quantitative agreement with our previous work~\cite{Nikolaev2024} for the same systems. This clearly demonstrates the power of the Wannierization method. Here, we propose further analyses emphasizing on the orbital torkance response related to our recent experiments~\cite{Krishnia2023}. 

In most cases the SOT exerted onto a ferromagnet within a HM/FM bilayer originate from two main sources: the spin/orbital Hall effect in the non-magnetic layer able to inject polarized electrons flowing normal to the HM/FM interface and giving rise to a net integrated spin torque along the $-\vec{M} \times (\vec{M} \times \delta \vec{s})$ direction, where $\vec{m}$ is the unit vector of the magnet and $\delta \vec{s}$ points along the out-of-equilibrium spin orientation generated by the spin-current. This scenario is usually related to the Berry curvature emerging from the HM bulk band structure. The other contribution arises from spin/orbital accumulation generated by inversion symmetry breaking. It is inherently correlated to the interface between the FM and NM or HM as a response to an electric field parallel to the plane such that it mainly points out along $\vec{M} \times \delta \vec{s}$, giving rise to a local torque of field-like torque (FLT) component. The current-driven field component can be heuristically computed as $\vec{h}_i=(\Delta_{xc}/\mathcal{V} M_s)\delta \vec{s}_i$, with $\Delta_{xc}$ the \textit{sp}-\textit{d} exchange potential between itinerant (NM) and localized (FM) electrons, $\delta \vec{s}_i$ is the $i$-th Pauli spin matrix in regular cartesian coordinates, $M_s$ is the saturation magnetization of the ferromagnet, $\mathcal{V}$ is the volume of the unit cell. 
The out-of-equilibrium quantity $\delta \vec{s}_i$ is computed within the linear response formalism after quantum statistical averaging considering the symmetrized decomposition of Kubo-Bastin formula proposed in Ref.~\cite{PhysRevB.102.085113}.

SOT is made possible via the SOI. However, at first glance the low-Z atoms constituting Co/Al bilayers is not expected to contribute much. Nevertheless, recent studies have pointed out the existence of large oREE in oxidized interfaces~\cite{otani2021,ding2022,otani2023,Krishnia2024,bony2025} and also at the interface between Co and metallic Al~\cite{Krishnia2023,Nikolaev2024} displaying unprecedented torques (FLT) with materials free of large SOI. We then conducted simulations of the SOT based on linear response theory. The full Hamiltonian can be decomposed into the kinetic part, $\hat{\mathcal{H}}_K$, a SOI term, $\hat{\mathcal{H}}_{SOC}=\xi_{SOC}\left(\hat{\nabla} V(r)\times \hat{p}\right)\cdot \hat{\mathbf{\sigma}}$, small but not zero for a 3\textit{d} magnet plus a magnetic or exchange part $\hat{\mathcal{H}}_{xc}=-\mu_B\mathbf{B}_{xc}\cdot\mathbf{\hat{\sigma}}$, where the exchange field $\mathbf{B}_{xc}$ accounts for the difference between the effective Kohn-Sham potentials for minority and majority quasi-particles in Co. By applying $\mathcal{E}$, an out of equilibrium change in the magnetization $\delta \mathbf{M}$ enables the modification of the exchange following $\delta \mathbf{B}_{xc}=B_{xc}\delta \mathbf{M}/M$. Such a dynamical effect leads to a torque $\mathcal{T}$ action onto $\textbf{M}$ within the unit cell given according to:

\begin{equation}
    \mathcal{T}=\int d^3r \mathbf{M} \times \delta \mathbf{B}_{xc}=-\int d^3r \mathbf{B}_{xc} \times \delta \mathbf{M}.
\end{equation}

As our approach is based on the linear response, the torque $\mathcal{T}$ can be written in terms of the so-called torkance $\mathbf{t}$ such that $\mathcal{T}=\mathbf{t}\mathcal{E}$~\cite{belashchenko1,belashchenko2}. The tensor $\mathbf{t}$ can be decomposed into a respective \textit{interband} and \textit{intraband} terms $\mathbf{t}_{ij}=\mathbf{t}^{\text{Inter}}_{ij}+\mathbf{t}^{\text{Intra}}_{ij}$, such that:

\begin{equation}
    \mathbf{t}_{ij}^{\text{Inter}}=\frac{e\hbar}{2\pi \mathcal{N}}\sum_{\mathbf{k},n\neq m}\frac{\operatorname{Im}\big [\braket{\psi_{\mathbf{k}n}|\mathcal{\hat{T}}_i|\psi_{\mathbf{k}m}}\braket{\psi_{\mathbf{k}m}|\hat{v}_j|\psi_{\mathbf{k}n}}\big]}{(\varepsilon_m-\varepsilon_n)^2}\nonumber
\end{equation}
and 
\begin{equation}
    \mathbf{t}_{ij}^{\text{Intra}}=\frac{e\hbar}{\pi \Gamma \mathcal{N}}\sum_{\mathbf{k},n} \operatorname{Re}\big [\braket{\psi_{\mathbf{k}n}|\mathcal{\hat{T}}_i|\psi_{\mathbf{k}n}}\braket{\psi_{\mathbf{k}n}|\hat{v}_j|\psi_{\mathbf{k}n}}\big] \delta(\varepsilon_F-\varepsilon_n)
\end{equation}
where we have adopted the clean limit case ($\Gamma\rightarrow 0$), but numerically is accounted for by a finite small value of the Fermi broadening energy $\Gamma=0.075$~meV (throughout the paper). $\varepsilon_F$ is the Fermi level, $\hat{\mathcal{T}}_i$ is the $i-$th component of the torque operator usually written as $\hat{\mathcal{T}}=\frac{-i}{\hbar}[\hat{\mathcal{H}},\hat{\mathbf{s}}]$ = $-\mu_B \hat{\mathbf{\sigma}} \times \mathbf{B}_{xc}$ with $v_j$ is the $j-$th component of the velocity operator. The exchange contribution to the Hamiltonian represents the main term owing in the case of a small SOI and is proportional to $\vec{\Omega}^{xc}=\Omega^{xc} \hat{m}$ and $\vec{\Omega}^{xc}=\frac{1}{2\mu_B}[V^{eff}_{min}(\vec{r})-V^{eff}_{maj}(\vec{r})]$ is the exchange field.  
We took this part as the values of individual atomic magnetic moments after the self-consistent calculation. From symmetry arguments, the partition into \textit{interband} and \textit{intraband} has to be assigned to mainly damping-like torque or DLT ($\mathcal{T}_{xy}$) and field-like torque or FLT ($\mathcal{T}_{xx}$) components respectively~\cite{Go_PRR2,Hayashi2023} despite the lack of a mirror symmetry due to the hexagonal stacking may induce small mixing terms between FL and DL terms~\cite{Nikolaev2024} (see also Table~\ref{table_xi} comparing the orbital polarization $\braket{\hat{l}_y}$ for respective \textit{intraband} vs. \textit{interband} terms).

\vspace{0.1in}

\begin{figure}[ht!]
    \centering
    \includegraphics[width=\linewidth]{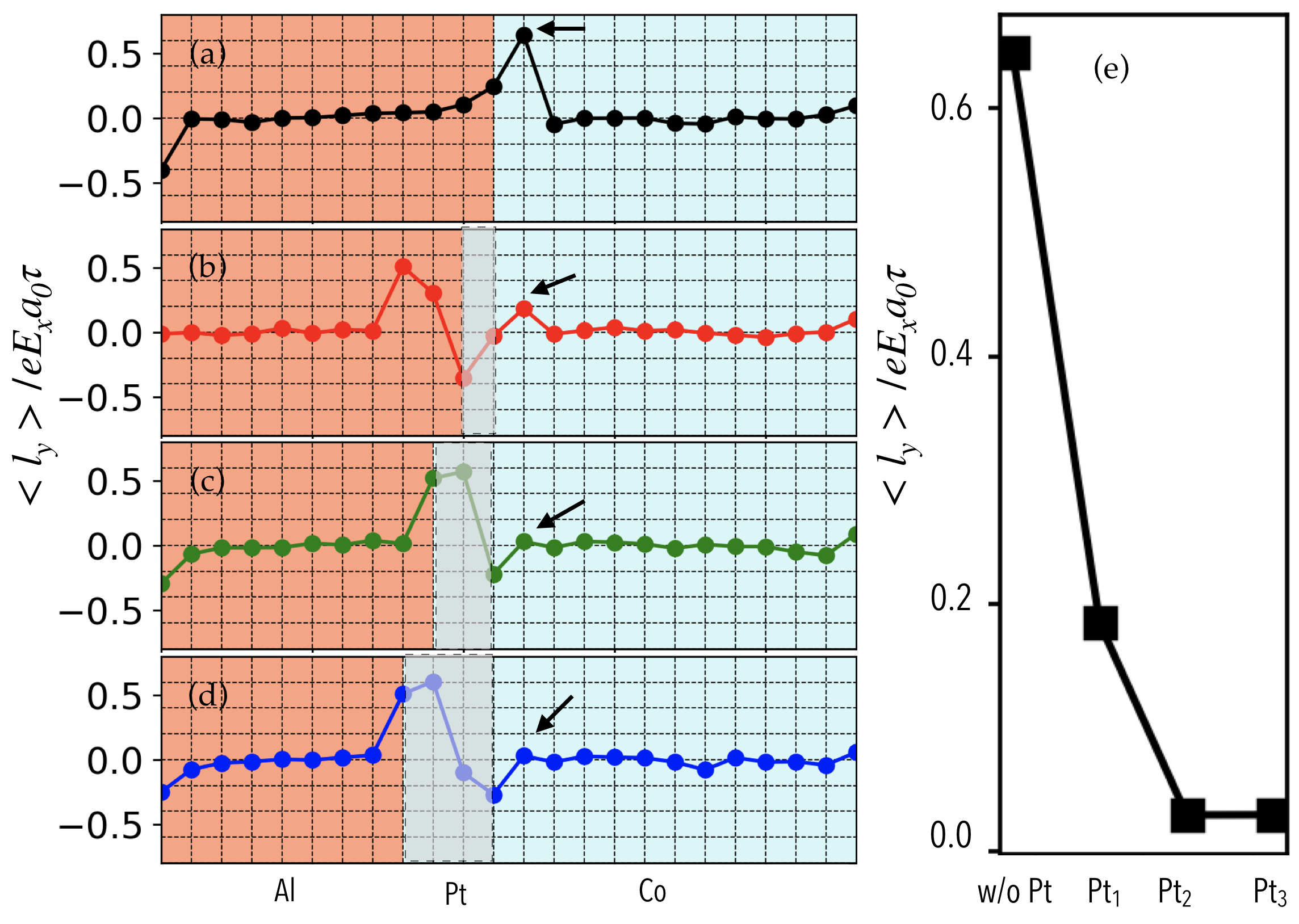}
    \caption{oREE calculated for Co(12)/Al(12) bilayers for a value of broadening energy $\Gamma=0.075$~eV (a) and with one (b), two (c) and three (d) Pt layers insertion. In (e) we show how the oREE on Co(1) signalized by black arrows from (a) to (d) decreases with the number of Pt layers.}
    \label{fig:OREE}
\end{figure}

\noindent
{\it Co/Al structures and orbital Rashba-Edelstein effects} - We turn to the evaluation of both spin and orbital accumulation and related torkance response from the determination of the inverse Rashba-Edelstein $\chi_{xy}^{REE}$ tensors. We also considered the layer-resolved torkance of the Pt(10)/Co(3) SHE reference leading to same qualitative results than previous published works~\cite{kirill_2022,freimuth_2014}. We retained the \textit{intraband} term~\cite{salemi2021} according to:

\begin{equation*}
    \braket{\hat{l}_y}=-\frac{e\hbar}{4\pi}\int \partial_{\epsilon}f_{\epsilon}d\epsilon \operatorname{Re} \operatorname{Tr} \bigg\{ \hat{l}_y (G_0^R-G_0^A)
    \hat{v}_x(G_0^R-G_0^A)\bigg\}\mathcal{E}_x,
\end{equation*}
for $\mathcal{E}_x$ along the $\hat{x}$ direction and show the orbital accumulation projected on each layer of the structure as depicted in Fig.~\ref{fig:OREE}. For Co(12)/Al(12), the response profile $\chi^{oREE}_{xy}=\frac{\hbar\braket{\hat{l}_y}}{e\mathcal{E}_xa_0\tau}$ displays a large peak of $\braket{\hat{l}_y}$ OAM density (in unit of $ea_0\tau \mathcal{E}$ with $a_0$ the Bohr radius and $\tau=\left(\frac{\hbar}{\Gamma}\right)$ the momentum or spin relaxation time) at the Co(1) atoms interfacing Al (Fig.~\ref{fig:OREE}(a)). We find an orbital response from the oREE tensor $\chi^{oREE}_{xy}$ dominant by almost two orders of magnitude compared to the equivalent spin term (\textcolor{blue}{Fig.~S9 SI-III}). Moreover, $\braket{\hat{l}_y}$ is shown to be almost suppressed upon the inclusion of a very few number of Pt planes (Pt1, Pt2, Pt3) sandwiched between Co and Al  (Fig.~\ref{fig:OREE}(e)). On adding Pt, we rather observe an oscillatory behavior of the generated OAM in Pt slightly propagating into neighboring Al layers (Fig.~\ref{fig:OREE}(b-d)). This feature may be understood as the spin-orbital imprint of the spin-polarization generated by sREE. Concerning Co(12)/Al(12), results are in very close agreement with Nikolaev et al.~\cite{Nikolaev2024} using \textit{Wannierization} techniques. The emergence of a very small spin-component (sREE) for Co/Al compared to the oREE counterpart (see table~\ref{table_xi}) accounts for a strongly reduced out-of-equilibrium spin accumulation $\delta \hat{s}(\hat{l})=\eta_{c-s(l)}\hat{z}\times \hat{j}_c$ unlike what is generally put forward for Rashba systems. This makes the strong peculiarity of such Co/Al system. 

\noindent We can give a fair estimate of the FLT component expected from oREE for Co/Al. Taking values for $\mathcal{E}_x \sim 2.5 \cdot 10^4$ V/m (equivalent to current density in Pt of $j_{Pt}=10^{11} A/$m$^2$) the OAM reaches values as large as $\chi_{xy}^{oREE} \approx 5.94 \times 10 ^{-10} \hbar$ $m/V/\text{atom}$  for which the effective FLT approaches $B_{FL}\approx\frac{\braket{\xi_{SO}}}{M_s t_F}\braket{l_y}=\frac{\braket{\xi_{SO}}}{M_s t_F }\chi_{xy}^{oREE}\mathcal{E}=2.7$~mT in pretty good agreement with our experiments~\cite{Krishnia2024} and theory and calculation considering a slightly different lattice parameter~\cite{Nikolaev2024}. We show that $\chi_{xy}^{oREE}$ may vary vs. the in-plane lattice parameter ($a$) or strain but keeping substantial values over a large range of $a$ (\textcolor{blue}{SI-III}).

\vspace{0.1in}

\begin{figure}[ht!]
\includegraphics[width=\linewidth]{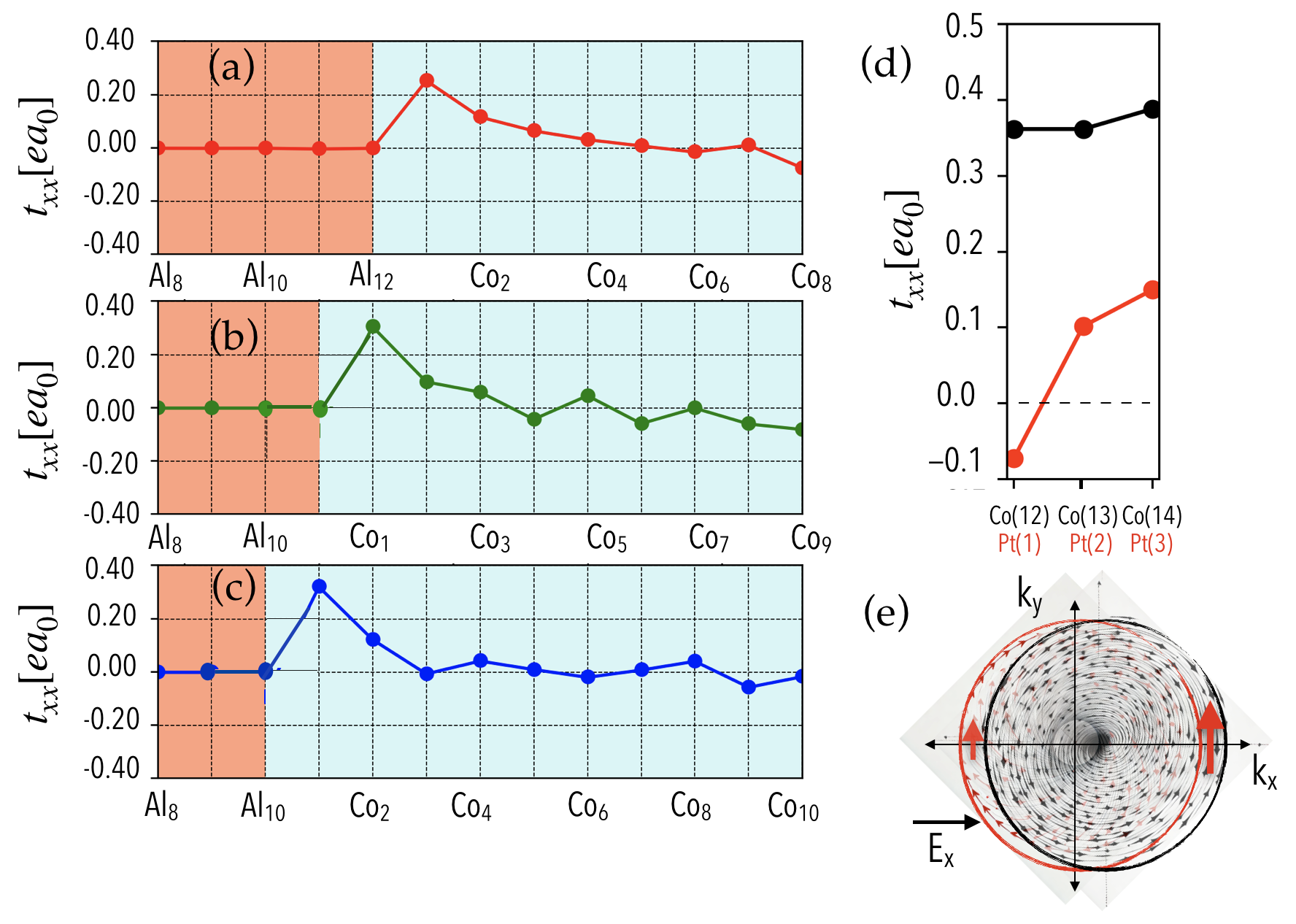}
\caption{Torkance components $t_{xx}$ (mainly FLT) in units $ea_0$ calculated for Co/Al bilayers with increasing numbers of FM layers, namely, (a) corresponds to the $t_{xx}$ for Al(12)/Co(12) bilayers, (b) Al(11)/Co(13) and (c) Al(10)/Co(14) bilayers respectively. In (d) we show the integrated value of t$_{xx}$ over four layers within the FM arising from the OREE schematically depicted in (e). }\label{fig:Torke_1}
\end{figure}

\noindent
{\it Torques at Co/Al interfaces:} We have explicitly calculated both the FL and DL torque components considering respective \textit{intraband} $t_{xx}$ (FLT) and \textit{interband} $t_{xy}$ (DLT) torkance. $t_{xx}$ is shown in Fig.~\ref{fig:Torke_1} for Co/Al as well as Co/Pt(1,2,3)/Al interfaces (see also Table~\ref{table_xi}). For Co/Al, we have chosen the following setups, a) Al(12)/Co(12), b) Al(11)/Co(13) and c) Al(10)/Co(14) respectively and equivalent structures for Pt(1,2,3) insertion.
The layer projected torkance depicted in Fig.~\ref{fig:Torke_1} (a,b,c) shows a gradual monotonous (non-oscillating) decaying behavior reaching its maximum at the interfacial Co(1) layer manifesting thus its strong orbital character~\cite{longrangeorbital}. On the Al side, although the orbital accumulation is not zero for Al(12) neighboring Co(1) (Fig.~\ref{fig:OREE}(a)), the torkance $t_{xx}$ is calculated to be strictly zero because of the absence of any induced magnetic moment on Al(12) unlike the case of Pt (see table~\ref{table:transport_coal}). We have also considered the cases of Fe(12)/Al(12) (Co being replaced by Fe) and Co$_{0.5}$Fe$_{0.5}$(6)/Al(6) for which the calculated torkances and orbital accumulation are depicted in table~\ref{table:transport_coal}). We note a strong decrease in the value of both orbital polarization and torkances upon Fe inclusion suggesting the peculiar role played by the Co/Al interface. From Ref.~\cite{Go_PRR2}, the \textit{intraband} torque $t_{xx}$ is more related to the spin-current influx ($\braket{\mathcal{Q}^{\mathbf{s}}}^{\text{intra}}(r)$) according to $\braket{\mathcal{T}_{xc}^{\mathbf{s}}}(r)^{\text{Intra}}=-\braket{\mathcal{T}_{SO}^{\mathbf{l}}}(r)^{\text{intra}}+\nabla_r \braket{\mathcal{Q}^{\mathbf{s}}}^{\text{intra}}(r)$, both quantities to the right being linearly linked to the SOI strength ($\xi_{SO}$). We have indeed checked that \textit{i}) the torkance $t_{xx}$ disappears when the SOI is switched off in both Co and Al, \textit{ii}) almost zero when SOI is absent in Co (only present in Al) or even absent only in Co(1) and \textit{iii}) almost unaffected when SOI is only absent in Al in agreement with out expectations. Integrated torkance on the FM side is shown in Fig.~\ref{fig:Torke_1}(d) where a very large FLT (black) takes place in the Co/Al systems (almost constant for the three structures) and the much smaller ones (red) for equivalent structures with Pt1, Pt2 and Pt3 insertion. An integrated value of the torkance of 0.37~$[ea_0]$ is equivalent to a field torque $B_{FL}=\left(\frac{e^*a_0}{ M_{s} N}\right)t_{xx}\mathcal{E}$ ($M_{s}$ is the atomic magnetic moment and $N$ the number of Co atomic planes) giving thus $B_{FL}\approx 2~$mT (for $\mathcal{E}=2.5\times 10^{4}$~V/m or $j_C^{Pt}=10^{11}~A/$m$^2$).

\begin{table}
\caption{\label{table:transport_coal} Co/Al and Co/Pt(1,2,3)/Al systems: $\braket{\hat{l}_y}$-orbital ($\chi_l^{oREE}$) and $\braket{\hat{\sigma}_y}$-spin ($\chi_s^{sREE}$) Rashba-Edelstein response at Co(1) interfacing Al or Pt. Values are given per atom. Integrated \textit{intraband} $t_{xx}$ (FL) and \textit{interband} $t_{xy}$ (DL) torkance components for Co/Al interfaces without/with Pt insertion.}
\begin{ruledtabular}
 \begin{tabular}{cccc} 
  Co/Al &$\chi_l^{intra}$$[10^{-10}\hbar$ $m/V]$ & $\chi_l^{inter}$$[10^{-10}\hbar$ $m/V]$& $t_{xx}$$[ea_0]$ \\ 
 \hline
 w/o Pt  & 5.94 & 0.20 &  0.37 \\ 
 1 Pt    & 1.56 & 0.18 & -0.08 \\
 2 Pt    & 0.52 & 0.22 &  0.10 \\
 3 Pt    & 0.44 & 0.16 &  0.15 \\ 
 Fe/Al   &  2.05  &  0.25 & 0.13  \\
 Co$_{0.5}$Fe$_{0.5}$/Al & 3.85 & 0.19 & 0.26 \\
 \hline
   &$\chi_s^{intra}$$[10^{-10}\hbar$ $m/V]$ & $\chi_s^{inter}$$[10^{-10}\hbar$ $m/V]$& $t_{xy}$$[ea_0]$ \\ 
 \hline
 w/o Pt  & 0.12 & -0.002  &  -0.04 \\ 
 1 Pt    & -0.3 & -0.01&  0.33 \\
 2 Pt    & -0.42 & -0.14 & -0.08 \\
 3 Pt    & -0.50 &  0.02 &  0.13 \\
  Fe/Al   &  0.18  &  0.07 & 0.05  \\
 Co$_{0.5}$Fe$_{0.5}$/Al & 0.38 & 0.08 & -0.05 \\
 \end{tabular}
 \end{ruledtabular}
 \label{table_xi}
\end{table}

\begin{figure}[ht!]
\includegraphics[width=\linewidth]{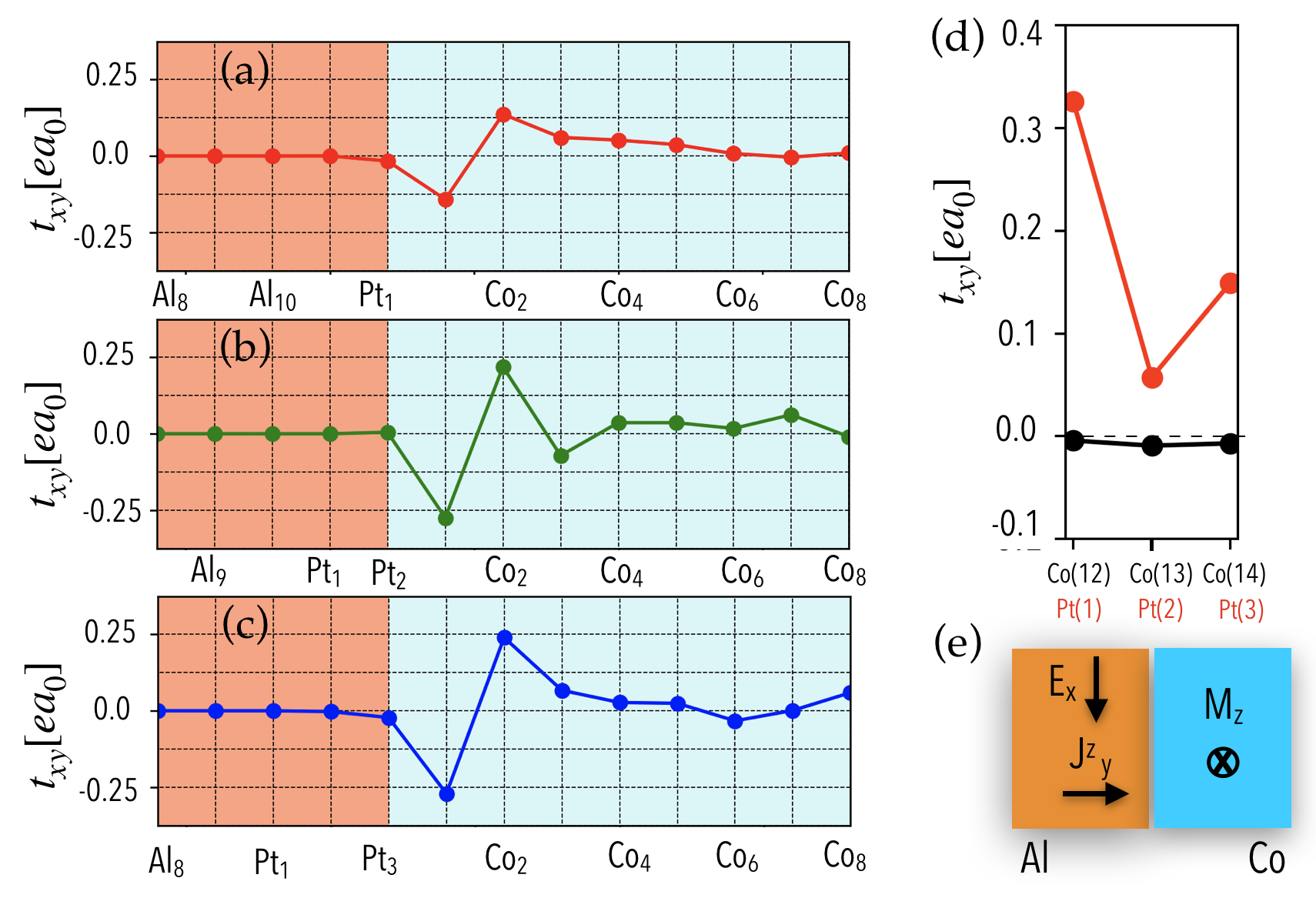}
\caption{$t_{xy}$ [$ea_0$] torkance (DLT) calculated for Al/Pt(1,2,3)/Co. (a) $t_{xy}$ for Al(11)/Pt(1)/Co(12) bilayers, (b) Al(10)/Pt(2)/Co(12) and (c) Al(9)/Pt(3)/Co(14) bilayers respectively. (d) Integrated value of t$_{xy}$ over four layers within the FM arising from the SHE schematically depicted in (e). Results for Co/Al in (d) showcasing almost no DLT (black).}\label{fig:Torke_2}
\end{figure}

\noindent The almost absence of oREE upon Pt atomic insertion deserves some comments. It is experimentally demonstrated (unpublished) that inserting Pt leads to enhancement of the DL component ($t_{xy}$) while strongly reducing the FLT ($t_{xx}$). To access this information, we consider the Fermi sea (\textit{interband}) component of the $t_{xy}$ torkance derived from the Kubo formula (table~\ref{table_xi} \textcolor{blue}{SI-III}). Those are displayed in Fig.~\ref{fig:Torke_2}, where $t_{xy}$ is layer projected for the case of Pt1(a), Pt2(b) and Pt3(c). We demonstrate quite large values for the DLT with integrated value of the $t_{xy}$ torkance depicted in Fig.~\ref{fig:Torke_2}(d) of the same order of magnitude than $t_{xx}$ previously discussed. Moreover, a clear oscillating character from plane to plane of the DLT (as for FLT) in the case of Pt insertion exhibits the exchange interactions between spin and local moment mainly arising from either spin-Hall effect (SHE) or sREE from Pt layer(s)~\cite{longrangeorbital}. We recover the SOT magnitude obtained on reference Co(3)/Pt(10) (\textcolor{blue}{SI-III}) samples together with a characteristic oscillating behavior are in close agreement with Refs.~\cite{belashchenko1,belashchenko2}.

\vspace{0.1in}

\noindent Coming back to Co(12)/Al(12), $t_{xy}$ values (DLT) remain small (Fig.~\ref{fig:Torke_2}(d)) indicating that both \textit{interband} orbital and spin torkance contribute much less than their \textit{intraband} orbital counterpart (FLT). This emphasizes the strong local character of the orbital accumulation on Co(1) plane preventing thus any momentum precession inside Co. This is made possible by the confinement experienced by the interfacial evanescent states. The relationship between the OAM generated by oREE and the electronic escape time into the bulk Co states has been discussed in a previous work~\cite{Fert2013}.

\vspace{0.1in}

\noindent {\it Discussion -} Our simulations reveal a significant difference between Co/Al and other type of interfaces (as for Co/Cu taken as a reference~\cite{Krishnia2023}).The Mulliken charge analysis indicates a nearly order-of-magnitude difference between the two systems, which accounts for the pronounced oREE observed at the Co/Al interface (\textcolor{blue}{SI-IV}). In comparison, the introduction of Pt resulted in a smaller Mulliken charge value relative to the Co/Al interface, underscoring the importance of chemical bonding among the atomic constituents. Furthermore, as listed in the \textcolor{blue}{SI-IV} and for direct comparison, we also have considered Co/Ti, Co/Mg interfaces giving negligible values for the OAM polarization as well as both FLT and DLT demonstrating thus the very specificity of Co/Al systems among other possible LM to observe oREE phenomenon. More generally, from our simulations we can establish a direct connection between the chemical properties of the FM/NM interface, the orbital Rashba effect and the associated orbital torques. We have extended our investigation to a new case, unaddressed before: the interface between Co and Ga of electronic configuration [\textit{Ar}]4\textit{s}$^2$3\textit{d}$^{10}$4\textit{p}$^1$ (\textcolor{blue}{SI-II}). Then, we can postulate that the electronic configuration of Al ([Ne]3\textit{s}$^2$3\textit{p}$^1$) is directly responsible for the appearance of the orbital Rashba effect near the K point. From a chemical standpoint, this distinctive configuration enables asymmetric hopping due to \textit{p–d} hybridization at interfaces. As we show below, this condition is not met in the other systems discussed in our manuscript. For example, the filled \textit{s} orbitals and partially filled \textit{d} states in Cu, Cr, and Ti do not lead to the effect at the Fermi level. 

To conclude, we have calculated the SOT for each odd and even orbital torque component, showing that our calculations based on the Kubo formalism ruled out the presence of a damping-like contribution for Co/Al. The observed decay on inserting Pt atomic planes, of mainly SHE character, aligns with the decreasing FL component, thereby validating our simulations. These results are significant because they enable us to study the anatomy of the oREE in the absence of a spin counterpart, which could guide the development of future spintronics devices with low-Z components. 

\vspace{0.1in}

\begin{acknowledgments}
The authors thank Albert Fert, Nicolas Reyren and S. Nikolaev (Univ. Osaka, Japan) for fruitful discussions. This study has been supported by the French National Research Agency under the project ANR-22-CE30-0026 'DYNTOP', by a France 2030 government grant managed by the French National Research Agency PEPR SPIN ANR-22-EXSP0009 (SPINTHEORY) and by the EIC Pathfinder OPEN grant 101129641 'OBELIX'. N.
Sebe benefits from a France 2030 government grant managed by the French National Research Agency (ANR-22-PEPR-0009 Electronique-EMCOM).
\end{acknowledgments}

\bibliography{bibibiography_tex,main}
\end{document}